\documentclass[conference]{IEEEtran}
\IEEEoverridecommandlockouts
\usepackage{cite}
\usepackage{amsmath,amssymb,amsfonts}
\usepackage{algorithmic}
\usepackage{graphicx}
\usepackage{textcomp}
\usepackage{xcolor}
\usepackage{hyperref}
\usepackage{booktabs}
\usepackage{braket}
\usepackage{comment}
\usepackage{bm}
\usepackage{multirow}
\usepackage{hyperref}
\usepackage{orcidlink}
\def\BibTeX{{\rm B\kern-.05em{\sc i\kern-.025em b}\kern-.08em
    T\kern-.1667em\lower.7ex\hbox{E}\kern-.125emX}}

\begin{document}

\title{Emergent Problem-Graph Alignment in RL-Discovered Entanglement Topologies for QAOA}

\author{
    \IEEEauthorblockN{
        Tobias Rohe\IEEEauthorrefmark{1}\IEEEauthorrefmark{4}\orcidlink{0009-0003-3283-0586}, 
        Federico Harjes Ruiloba\IEEEauthorrefmark{1}\IEEEauthorrefmark{2}\orcidlink{0009-0002-2283-0921}, 
        Markus Baumann\IEEEauthorrefmark{1}\orcidlink{0009-0007-3575-1006}, 
        Gerhard Stenzel\IEEEauthorrefmark{1}\orcidlink{0009-0009-0280-4911}, \\
        Leo Sünkel\IEEEauthorrefmark{1}\orcidlink{0009-0001-3338-7681}, 
        Thomas Gabor\IEEEauthorrefmark{3}\orcidlink{0000-0003-2048-8667} and 
        Claudia Linnhoff-Popien\IEEEauthorrefmark{1}\orcidlink{0000-0001-6284-9286}
    }
    \IEEEauthorblockA{
        \IEEEauthorrefmark{1}Institute for Computer Science, LMU Munich, Munich, Germany\\
    }
    \IEEEauthorblockA{
        \IEEEauthorrefmark{2}relAI – Konrad Zuse School of Excellence in Reliable AI, LMU Munich, Munich, Germany
    }
    \IEEEauthorblockA{
        \IEEEauthorrefmark{3}Department of Computer Science, University of Exeter, Exeter, United Kingdom
    }
    \IEEEauthorblockA{
        \IEEEauthorrefmark{4}Email: tobias.rohe@ifi.lmu.de
    }
}

\maketitle

\begin{abstract}
In the Quantum Approximate Optimization Algorithm (QAOA), the entanglement topology, where qubit pairs are connected by two-qubit gates, is conventionally set equal to the edge set of the problem graph. This coupling ties circuit design to explicit problem knowledge and may not yield the most trainable circuit under limited optimization budgets. We investigate whether a reinforcement learning (RL) agent can discover more effective entanglement topologies for QAOA-based MaxCut optimization without direct access to the problem graph. A Masked Proximal Policy Optimization agent sequentially places IsingZZ gates to construct a circuit topology, while a variational inner loop optimizes the resulting QAOA parameters and returns the approximation ratio as a sparse terminal reward. The agent's observation contains only the edges placed so far and the current approximation ratio; graph structure can only be inferred indirectly through the optimization reward. On Erd\H{o}s--R\'{e}nyi instances with up to $10$~qubits, the agent consistently converges to topologies that are strict subsets of the problem graph, achieving overlap ratios approaching $1.0$, despite receiving no explicit information about the graph structure in its observations. These sparse, problem-aligned topologies outperform the full graph topology and several structural baselines when the optimization budget is limited ($50$~gradient steps), but are overtaken by denser topologies given sufficient optimization budget. Our results reveal a trainability--expressibility trade-off governed by topology density and suggest that the variational optimization landscape implicitly encodes structural information about the problem Hamiltonian.
\end{abstract}


\section{Introduction}
\label{sec:introduction}
Variational quantum algorithms (VQAs) are among the most actively studied approaches for leveraging near-term quantum hardware~\cite{cerezo2021variational, bharti2022noisy}. By combining parameterized quantum circuits with classical optimization, VQAs can approximate solutions to problems in chemistry, machine learning, and combinatorial optimization without requiring the deep, fault-tolerant circuits assumed by many quantum algorithms. The Quantum Approximate Optimization Algorithm (QAOA)~\cite{farhi2014quantum} is a prominent instance of this paradigm, designed specifically for combinatorial problems such as \textsc{MaxCut}.
 
A central challenge in applying VQAs is the design of the circuit ansatz: which gates to include, how to arrange them, and---critically---which qubit pairs to connect with entangling operations. This last choice, the \emph{entanglement topology}, determines which correlations the circuit can express and has a direct impact on both solution quality and trainability~\cite{sim2019expressibility}. In standard QAOA, the topology is fixed to the edge set of the problem graph, ensuring that each term in the cost Hamiltonian is directly addressed by a corresponding gate. While natural, this convention couples the circuit structure to explicit problem knowledge.
 
Several recent works have begun to treat circuit structure as a learnable quantity rather than a fixed design choice. Reinforcement learning (RL) and evolutionary methods have been applied to gate selection and ansatz construction for VQE~\cite{Ostaszewski2021ReinforcementLFA, Grimsley2018AnAVC, Huang2022RobustRQE, wang2022quantumnas}, and entanglement layout search has been explored for quantum machine learning~\cite{Nguyen2021QuantumESG}. However, the specific question of learning entanglement topologies for QAOA in the combinatorial optimization setting and analyzing \emph{what} the learned topologies reveal about the relationship between entanglement structure and problem structure has received limited attention. On the one hand, this is of course due to the naturally derived circuit architecture from the cost Hamiltonian and its associated advantages; however, the question certainly arises as to whether heuristic circuits with potentially different properties—specifically, different cost layers—can also be developed. 
 
In this work, we investigate whether an RL agent, guided solely by the approximation ratio obtained from a variational inner loop, can discover effective entanglement topologies for QAOA-based MaxCut optimization. Our framework consists of two nested loops: an outer loop in which a Masked Proximal Policy Optimization (PPO) agent sequentially places IsingZZ entangling gates, and an inner loop in which an Adam-based optimizer trains the resulting QAOA circuit on a given MaxCut instance. Crucially, the agent's observation space contains no explicit information about the problem graph; it can only observe the edges it has placed so far and the resulting approximation ratio.
While the agent interacts with the same problem instance across all episodes of a training run, it receives no direct encoding of the graph edges---any structural knowledge must be inferred indirectly through the variational optimization reward.
This design isolates the question of whether problem-relevant topology information is encoded in the optimization landscape and can be extracted through reward maximization.
 
We evaluate the framework on unweighted Erd\H{o}s--R\'{e}nyi random graphs with up to $10$~qubits and report three main findings:
 
\begin{enumerate}
    \item \textbf{Emergent problem-graph alignment.} The RL agent consistently converges to topologies that are strict subsets of the problem graph. On $8$- and $10$-qubit instances, the overlap ratio (the fraction of agent-selected edges that belong to the problem graph) reaches $\approx 1.0$ across all seeds, even though the agent has no direct access to the graph structure in its observations. This indicates that the variational optimization landscape encodes structural information about the cost Hamiltonian, which the agent discovers through reward maximization.
 
    \item \textbf{Trainability--expressibility trade-off.} The sparse, learned topologies outperform the full problem-graph topology and several structural baselines (linear, ring, random subsets) when the optimization budget is limited to $50$~gradient steps. However, with sufficient budget ($500$~steps), denser topologies recover and eventually surpass the RL topology. This reveals that the topology's primary role, in the regimes studied, is to control convergence speed rather than asymptotic solution quality.
 
    \item \textbf{Entanglement structure analysis.} Using the Meyer--Wallach entanglement measure, we characterize the global entanglement of the output states produced by different topologies, providing additional evidence that the learned topologies produce entanglement profiles tailored to the optimization task.
\end{enumerate}
 
The remainder of this paper is organized as follows. Section~\ref{sec:background} provides background on MaxCut, QAOA, circuit topology, and reinforcement learning. Section~\ref{sec:relatedwork} reviews related work on automated circuit design. Section~\ref{sec:methods} describes our MDP formulation, QAOA ansatz, and training procedure. Section~\ref{sec:experimental_setup} details the experimental setup, and Section~\ref{sec:results} presents our results. We discuss implications and limitations in Section~\ref{sec:discussion} and conclude in Section~\ref{sec:conclusion}.

\section{Background}
\label{sec:background}

\subsection{MaxCut and the Ising Formulation}
\label{sec:maxcut}

Given an undirected graph $G = (V, E)$ with $|V| = n$ nodes and optional edge weights $w_{ij} > 0$, the \textsc{MaxCut} problem asks for a bipartition of $V$ into two disjoint sets $S$ and $\bar{S}$ that maximizes the total weight of edges crossing the partition.
This problem is NP-hard in general~\cite{Karp1972Reducibility} and admits a direct mapping to an Ising Hamiltonian:
\begin{equation}
  H_C = -\sum_{(i,j) \in E} w_{ij}\,\frac{1 - Z_i Z_j}{2}\,,
  \label{eq:maxcut_hamiltonian}
\end{equation}
where $Z_i$ denotes the Pauli-$Z$ operator on qubit~$i$.
Each computational-basis state $|z\rangle$ encodes a candidate partition, and the ground state of $H_C$ corresponds to a maximum cut.
The direct Ising mapping and scalable hardness make \textsc{MaxCut} a standard benchmark for variational quantum optimization.

We quantify solution quality via the \emph{approximation ratio}
\begin{equation}
    \mathrm{AR} = \frac{-\langle \psi | H_C | \psi \rangle}{C^*}\,,
  \label{eq:ar}
\end{equation}
where $C^* > 0$ is the optimal cut value (i.e., the maximum total weight of edges crossing any partition).
The negation converts the ground-state energy of~$H_C$ (which is negative) to a positive quantity.

An $\mathrm{AR}$ of~$1$ indicates an optimal solution, while the uniform superposition yields in expectation $\mathrm{AR} \approx 0.5$ on typical instances.

\subsection{The Quantum Approximate Optimization Algorithm}
\label{sec:qaoa}

The Quantum Approximate Optimization Algorithm (QAOA)~\cite{farhi2014quantum} is a variational framework tailored to combinatorial optimization.
A QAOA circuit of depth~$p$ alternates two parameterized unitaries across $p$~layers.
Starting from the uniform superposition $|+\rangle^{\otimes n}$, layer~$l$ applies:
\begin{equation}
  |\boldsymbol{\gamma},\boldsymbol{\beta}\rangle
  = \prod_{l=1}^{p}\;
    U_M(\beta_l)\;U_C(\gamma_l)\;|+\rangle^{\otimes n}\,,
  \label{eq:qaoa_ansatz}
\end{equation}
where the \emph{cost unitary} $U_C(\gamma_l) = \prod_{(i,j)\in T} \mathrm{IsingZZ}_{ij}(\gamma_l)$ imprints the problem structure, and the \emph{mixer unitary} $U_M(\beta_l) = \prod_{q=1}^{n} R_X^{(q)}(\beta_l)$ drives exploration across partitions.
In standard QAOA, the gate topology~$T$ is identical to the problem edge set~$E$.
A classical optimizer—typically gradient-based (e.g., Adam with parameter-shift gradients)—then minimizes $\langle H_C \rangle$ over $(\boldsymbol{\gamma}, \boldsymbol{\beta})$.

Parameter initialization significantly affects convergence.
The \emph{ramp schedule}~\cite{zhou2020quantum} initializes $\gamma_l$ and $\beta_l$ as linearly interpolated values across layers, providing a structured starting point that outperforms random initialization in practice.

Crucially, standard QAOA \emph{couples} the circuit topology to the problem graph: gates are placed exactly on the edges of~$E$.
Our work decouples this relationship and treats the topology as an independent design variable discovered by a learning agent.

\subsection{Circuit Topology as a Design Variable}
\label{sec:topology}

We define the \emph{circuit topology} $T \subseteq \binom{[n]}{2}$ as the set of qubit pairs on which entangling gates act.
In our QAOA design, topology becomes the decisive factor:
since $|+\rangle$ is an eigenstate of~$X$, the $R_X$ mixer leaves it invariant, and the circuit can only depart from the uniform superposition through $\mathrm{IsingZZ}$ gates.
With $T = \emptyset$, no $Z$-$Z$ correlations are generated and $\mathrm{AR} \approx 0.5$ regardless of the parameters.
This stands in contrast to hardware-efficient ans\"atze for VQE, where single-qubit rotations preceding entangling layers can already produce arbitrary product states~\cite{kandala2017hardware}; there, the entangling topology is not the sole source of expressivity.

Given a problem graph~$G = (V, E)$, several canonical topology choices exist: the \emph{graph-aligned} topology ($T = E$), \emph{linear} nearest-neighbor chains, \emph{ring} topologies, and random subsets of varying density.

As a complementary metric, the \emph{Meyer--Wallach entanglement measure}~\cite{meyer2002global}
\begin{equation}
  Q(|\psi\rangle) = \frac{2}{n}\sum_{k=1}^{n}\left(1 - \mathrm{tr}\!\left(\rho_k^2\right)\right)
  \label{eq:mw}
\end{equation}
quantifies the average entanglement across all qubits, where $\rho_k$ is the reduced density matrix of qubit~$k$.
$Q = 0$ for product states and $Q = 1$ for maximally entangled states, providing a scalar summary of global entanglement in the output state.

\subsection{Reinforcement Learning and Masked Policy Optimization}
\label{sec:rl}
A Markov Decision Process (MDP) is defined by the tuple $(S, A, P, R, \gamma)$ comprising states, actions, transition dynamics, a reward function, and a discount factor~\cite{sutton1998reinforcement}.
At each timestep~$t$, an agent observes state $s_t \in S$, selects action $a_t \in A$ according to a policy $\pi(a|s)$, receives reward $r_t = R(s_t, a_t)$, and transitions to~$s_{t+1}$.
The objective is to find a policy that maximizes the expected discounted return $G_t = \sum_{k=0}^{\infty} \gamma^k r_{t+k}$.
MDPs provide a natural framework for sequential construction tasks where each decision incrementally builds a discrete structure.

Policy gradient methods parameterize the policy as $\pi_\theta(a|s)$ and optimize it by ascending the gradient of expected return~\cite{williams1992simple}.
Unlike value-based methods such as DQN, which learn a $Q$-function and derive a policy indirectly, policy gradient approaches optimize the policy directly.
This is particularly advantageous in settings with large discrete action spaces, where maintaining accurate value estimates for every state--action pair becomes impractical.

Proximal Policy Optimization (PPO)~\cite{schulman2017proximal} addresses a key instability of vanilla policy gradients: large parameter updates can cause significant performance drops.
PPO constrains the update magnitude via a clipped surrogate objective:
\begin{equation}
  L^{\mathrm{CLIP}}(\theta) = \hat{\mathbb{E}}_t\!\left[
    \min\!\left(r_t(\theta)\,\hat{A}_t,\;
    \mathrm{clip}\!\left(r_t(\theta), 1{-}\epsilon, 1{+}\epsilon\right)\hat{A}_t\right)
  \right],
  \label{eq:ppo}
\end{equation}
where $r_t(\theta) = \pi_\theta(a_t|s_t)/\pi_{\theta_{\mathrm{old}}}(a_t|s_t)$ is the probability ratio between the updated and previous policy, and $\hat{A}_t$ is the estimated advantage, typically computed via Generalized Advantage Estimation (GAE)~\cite{schulman2015high}.
When the ratio~$r_t$ deviates beyond $[1{-}\epsilon,\, 1{+}\epsilon]$, the objective is clipped, preventing the optimizer from exploiting a single large gradient step.
An auxiliary entropy bonus $\mathcal{H}[\pi_\theta]$ is commonly added to the objective to encourage exploration and prevent premature convergence to deterministic policies.
Together, these mechanisms make PPO robust to noisy reward signals, a property that is critical when rewards are derived from stochastic or computationally expensive evaluation procedures.

In many combinatorial construction tasks, the set of valid actions is not fixed but shrinks as decisions accumulate—for instance, an edge that has already been placed should not be selected again.
\emph{Action masking}~\cite{huang2022closer} addresses this by setting the logits of invalid actions to $-\infty$ before the softmax:
\begin{equation}
  \tilde{\ell}_i =
  \begin{cases}
    \ell_i & \text{if } a_i \text{ is valid,} \\
    -\infty & \text{otherwise,}
  \end{cases}
  \label{eq:masking}
\end{equation}
where $\ell_i$ denotes the raw logit for action~$a_i$.
After softmax, masked actions receive exactly zero probability, and the remaining probability mass is redistributed among feasible actions.
This is preferable to post-hoc penalty schemes (e.g., negative rewards for invalid actions), which still sample invalid transitions and distort the reward signal.
Action masking has been shown to significantly improve sample efficiency in constrained combinatorial environments~\cite{huang2022closer}.

Reinforcement learning is a natural fit for circuit architecture search: the sequential placement of entangling gates constitutes a discrete, step-by-step decision process over a combinatorial space that grows as $\mathcal{O}\!\left(\binom{n}{2}^{m_{\max}}\right)$, rendering exhaustive search infeasible even for moderate qubit counts.
Prior work has demonstrated the effectiveness of RL for quantum circuit synthesis and architecture discovery, motivating its application to the topology selection problem studied in this paper.

\section{Related Work}
\label{sec:relatedwork}
The design of variational quantum circuits, particularly the automatic optimization of their structure and entanglement patterns, has become a central research topic in the quest for quantum advantage on near-term devices. Recent literature explores a variety of approaches, including reinforcement learning, evolutionary algorithms, differentiable programming, and entanglement-focused search, to automate and enhance quantum circuit design for applications such as VQE and QAOA.

\textbf{Reinforcement Learning and Adaptive Ansatz Construction.}
Several works have leveraged RL and adaptive strategies to automate the construction of variational ans\"atze. Grimsley et al.~\cite{Grimsley2018AnAVC} introduce ADAPT-VQE, which incrementally appends individual operators to the ansatz based on gradient information from the target Hamiltonian, rather than committing to a predetermined circuit structure. This system-driven growth yields compact circuits that achieve chemical accuracy with substantially fewer parameters than standard unitary coupled cluster ans\"atze, including for strongly correlated molecular systems.

Ostaszewski et al.~\cite{Ostaszewski2021ReinforcementLFA} frame VQE ansatz construction as a deep RL problem in which an agent sequentially appends gates to a circuit, guided by a feedback-driven curriculum that progressively tightens the energy threshold.
Evaluated on the lithium hydride ground-state energy benchmark, the method reaches chemical accuracy while producing circuits that are considerably shallower than both hardware-efficient and UCCSD ans\"atze.

Sadhu et al.~\cite{Sadhu2024AQIJ} complement these methods with a quantum information-theoretic perspective on RL-driven QAS, analyzing concurrence bounds of the circuits proposed by the agent for the variational quantum
state diagonalization problem. Their analysis reveals a phase transition in the correlation structure of these bounds as input-state entanglement increases, and shows that pre-conditioning the input state with an entanglement-enhancing block doubles the agent's success rate while reducing circuit depth.

\textbf{Evolutionary and Differentiable Methods.}
Huang et al.~\cite{Huang2022RobustRQE} take an evolutionary approach to circuit design, introducing the QCEAT algorithm that encodes quantum circuits as variable-length genomes and evolves both their structure and gate parameters through mutation and crossover operations. Unlike fixed-topology designs such as the hardware-efficient ansatz, QCEAT makes no assumptions about circuit depth or layout, and the resulting circuits use significantly fewer gates---particularly two-qubit entangling gates---yielding improved robustness against both coherent and incoherent noise on hydrogen, water, and Heisenberg model Hamiltonians.

On the differentiable and noise-aware side, Wang et al.~\cite{wang2022quantumnas} propose QuantumNAS, which jointly searches over variational circuit architectures and physical qubit mappings while explicitly accounting for device noise. A shared SuperCircuit is trained once and then used to cheaply rank candidate sub-circuits, which are refined through evolutionary selection and iterative gate pruning. Evaluated on $14$ quantum devices across both classification and VQE tasks, the framework consistently outperforms fixed-topology baselines, underscoring the value of hardware-aware circuit optimization.

\textbf{Entanglement Structure Selection.}
Closest to the topology aspect of our work, Nguyen and Chen~\cite{Nguyen2021QuantumESG} address the problem of selecting which qubit pairs to entangle in parameterized circuits for quantum machine learning. Their QES algorithm represents CNOT-based entanglement patterns as directed multigraphs, encodes candidate layouts as genotype vectors, and searches over this space using surrogate-assisted sequential model-based optimization. On several classification benchmarks the automatically discovered layouts outperform hand-crafted entanglement patterns, confirming that the choice of entanglement topology materially affects circuit performance, an observation that motivates our own investigation in the combinatorial optimization setting.

\textbf{Positioning of the present work.}
While the methods above address ansatz construction for VQA or entanglement layout search for quantum machine learning, our work targets a distinct and complementary question: what entanglement topologies does an RL agent discover for \emph{QAOA-based combinatorial optimization}, and what do these topologies reveal about the relationship between circuit structure and problem structure?
Unlike prior RL-based approaches that optimize full gate sequences, we restrict the agent to selecting only the \emph{entanglement edges} while keeping the gate type (IsingZZ) and circuit template (QAOA) fixed, isolating the role of topology.
Crucially, our framework deliberately withholds all explicit problem-graph information from the agent's observation space: the agent observes only the edges it has placed and the resulting approximation ratio.
This information asymmetry turns the RL agent into a diagnostic probe of the variational optimization landscape.
The emergent alignment between learned topologies and the problem graph therefore constitutes evidence that the landscape itself encodes structural information about the cost Hamiltonian, a finding about the \emph{nature} of variational quantum optimization rather than merely its performance.
An overall performance gain is only secondary for this work.

\section{Methods}
\label{sec:methods}
We decompose the problem into two nested loops: an \emph{outer loop} where an RL agent constructs entanglement topologies, and an \emph{inner loop} where a VQA optimizer evaluates each topology by training a parameterized QAOA circuit on the MaxCut instance.

\subsection{MDP Formulation}
\label{sec:mdp}

We formulate entanglement topology discovery as a Markov Decision Process. 
At each step, the agent either adds one undirected edge to the topology or, alternatively, terminates the episode by choosing a special \texttt{STOP} action. After a maximal number of gates have been added or the \texttt{STOP} action has been selected by the agent, the resulting VQA is optimized for a fixed number of steps. The achieved approximation ratio after the inner optimization process serves as a reward for the agent.

Formally, we define the MDP as a tuple $(\mathcal{S}, \mathcal{A}, P, R, \gamma)$ with the following components:

\textbf{State space $\mathcal{S}$.}
The observation $s_t \in \mathbb{R}^{2m_{\max}+2}$ is a fixed-length real vector:
\begin{equation}
    s_t = [\;\underbrace{\mathbf{e}_{\text{seq}}}_{2 m_{\max}}\;,\;\; \underbrace{t / m_{\max}\vphantom{\mathbf{e}}}_{1}\;,\;\; \underbrace{\text{AR}_t\vphantom{\mathbf{e}}}_{1}\;].
\end{equation}
The vector is composed of three parts.

\emph{(i) Edge sequence} (first $2 m_{\max}$ entries, where $m_{\max} = 2n$ is the edge budget).
This region has $m_{\max}$ consecutive \emph{slots} of two floats each; slot $i$ stores the two qubit indices of the $i$-th edge added by the agent.
If $k$ edges have been placed so far, slots $1$ through $k$ are filled and slots $k{+}1$ through $m_{\max}$ remain at zero.
Qubit indices are normalized: qubit $q$ is encoded as $(q+1)/(n+1)$, mapping to the open interval $(0,1)$.
The value $0$ is therefore reserved as an ``empty'' sentinel that cannot represent any real qubit, allowing the network to distinguish occupied from vacant slots.

\emph{(ii) Step fraction} $t / m_{\max} \in [0,1]$, indicating how far the episode has progressed relative to the budget.

\emph{(iii) Current approximation ratio} $\text{AR}_t \in [0,1]$, the approximation ratio at the final iteration of the most recent VQA optimization. Initialized to $0$ at episode start.

For $n=10$ qubits the observation space has $2 \times 20 + 2 = 42$ dimensions.

\textbf{Action space $\mathcal{A}$ and masking.}
The agent selects from $\binom{n}{2} + 1$ discrete actions: one for each possible undirected qubit pair $(i,j)$ with $i < j$, plus a \texttt{STOP} action.
As the IsingZZ entanglement gate used in our QAOA experiments is symmetric, undirected pairs are sufficient. For $n=10$, there are 45 possible edge actions, resulting in an action space with 46 dimensions when including the \texttt{STOP} action.
A boolean \emph{action mask} $m_t \in \{0,1\}^{|\mathcal{A}|}$ is computed at each step:
\begin{equation}
    m_t(a) = \begin{cases}
        0 & \text{if } a \text{ is an edge already in } \mathcal{T}_t, \\
        1 & \text{otherwise,}
    \end{cases}
\end{equation}
where $\mathcal{T}_t$ denotes the current entanglement topology of the circuit.

The \texttt{STOP} action is always unmasked ($m_t(\text{\texttt{STOP}}) = 1$).
The mask is applied \emph{before} the policy softmax, ensuring zero probability for invalid actions during both action sampling and policy gradient computation.

\textbf{Reward.}
We use a \emph{sparse terminal reward}: the VQA inner loop is executed only when the agent terminates the episode (via \texttt{STOP} or upon reaching the edge budget $m_{\max}$), and the reward equals the final approximation ratio:
\begin{equation}
    R = \text{AR}_{\text{terminal}} = \frac{-\langle\psi(\bm{\theta}^*)|H_C|\psi(\bm{\theta}^*)\rangle}{C^*},
\end{equation}
where $\bm{\theta}^*$ are the VQA-optimized circuit parameters and $C^*$ is the brute-force maximum cut value.
Intermediate steps receive zero reward.
We also support a \emph{dense} reward mode. In this case, we optimize the VQA after every step, and compute $R_t$ as $\text{AR}_\Delta$, i.e. the difference between the AR at timestep $t$ and the AR from the previous step $t-1$. As the sparse mode was computationally cheaper and produced higher final ARs in general, we focus on this variant.

\textbf{Episode dynamics.}
At the start of each training run, an Erd\H{o}s--R\'{e}nyi graph $G(n, 0.5)$ is sampled and verified to be connected (every node has at least one edge); this graph remains fixed across all episodes within the run.
Each episode begins with a fresh random seed for VQA parameter initialization.
The agent starts with an empty topology $\mathcal{T}_0 = \emptyset$ and incrementally builds it.
The episode terminates when the agent selects \texttt{STOP} or the number of added edges reaches $m_{\max}$.
Crucially, the agent's observation contains \emph{no explicit information} about which edges belong to the problem graph---it can only observe the edges it has placed and the resulting VQA performance.
Any knowledge of graph structure must therefore be acquired indirectly through the optimization reward over the course of training.

\subsection{QAOA Ansatz}
\label{sec:ansatz}

Given an entanglement topology $\mathcal{T}$, as discovered by the RL agent, we construct a QAOA circuit with $p$ layers as follows.
Unlike standard QAOA, which uses a single $\gamma_l$ per layer shared across all cost gates, we assign an \emph{independent} variational parameter to each edge and each qubit per layer, yielding an over-parameterized ansatz that gives the optimizer finer-grained control. This per-gate parameterization follows the multi-angle QAOA (ma-QAOA) formulation of Herrman et al.~\cite{herrman2022multi}, which has been shown to lower-bound standard QAOA performance and to match the approximation quality of three conventional QAOA layers with a single ma-QAOA layer on MaxCut instances. This property is particularly relevant in our setting, where the computational cost of the VQA inner loop limits RL training to $p{=}1$ layers, making it essential to extract maximum expressivity from a single layer.

All qubits are initialized to $\ket{+} = H\ket{0}$ via Hadamard gates.
Each layer $l \in \{1, \ldots, p\}$ then applies:

\begin{enumerate}
    \item \textbf{Cost unitary:} An $\mathrm{IsingZZ}(\gamma_l^{(i,j)})$ gate on each edge $(i,j) \in \mathcal{T}$, implementing $e^{-i \gamma Z_i Z_j / 2}$.
    Each edge has its own variational parameter $\gamma_l^{(i,j)}$.
    \item \textbf{Mixer unitary:} An $R_X(\beta_l^q)$ rotation on each qubit $q \in \{0, \ldots, n-1\}$, implementing $e^{-i \beta X / 2}$.
\end{enumerate}

The full unitary is:
\begin{equation}
    U(\bm{\gamma}, \bm{\beta}) = \prod_{l=1}^{p} \left[\prod_{q=0}^{n-1} R_X(\beta_l^q) \prod_{(i,j) \in \mathcal{T}} \mathrm{IsingZZ}(\gamma_l^{(i,j)})\right].
\end{equation}
The total number of variational parameters is $p \times (|\mathcal{T}| + n)$.
For $p=1$ and a topology with 8 edges on a 10-qubit system, this is $1 \times (8 + 10) = 18$ parameters.

The $\ket{+}^{\otimes n}$ initialization, combined with $R_X$ mixers, is essential for making the topology meaningful for our experiments.
Since $\ket{+}$ is an eigenstate of $X$, the mixer alone cannot drive the state towards $Z$-basis product states.
Only the IsingZZ entangling gates on the topology edges create the $Z$-$Z$ correlations needed to reach computational basis states that encode MaxCut solutions.
This design choice ensures that the entanglement topology is a genuine performance bottleneck: with an empty topology ($\mathcal{T} = \emptyset$), the circuit can only produce the uniform superposition, yielding $\mathrm{AR} \approx 0.5$ for random graphs.

In all our experiments, QAOA parameters are initialized using a ramp schedule~\cite{zhou2020quantum}: $\gamma_l = \frac{l+1}{p+1} \cdot \frac{\pi}{2}$ and $\beta_l = (1 - \frac{l+1}{p+1}) \cdot \frac{\pi}{2}$, with small Gaussian perturbations ($\sigma = 0.01$) to break symmetry.

\subsection{VQA Inner Loop}
\label{sec:vqa}

The inner loop optimizes the QAOA circuit parameters $\bm{\theta} = (\bm{\gamma}, \bm{\beta})$ to minimize the expectation value of the MaxCut Hamiltonian (Eq.~\ref{eq:maxcut_hamiltonian}).
We use PennyLane's~\cite{bergholm2018pennylane} \texttt{default.qubit} statevector simulator with automatic differentiation via the parameter-shift rule, and the Adam optimizer~\cite{kingma2015adam} with learning rate $\eta = 0.05$.

During RL training, we run $50$ VQA optimization steps per episode to balance reward signal quality against computational cost.
For the post-training topology comparison (Section~\ref{sec:comparison}), we reuse $50$ steps for the tuned suite and $500$ steps for the baseline suite, each with $5$ independent random seeds per topology, matching the stored comparison artifacts.
The maximum cut value $C^*$ is computed exactly via brute-force enumeration over all $2^n$ bitstrings, which is feasible for $n \leq 10$.

\subsection{Masked Proximal Policy Optimization}
\label{sec:ppo}

We train the RL agent using \emph{Masked Proximal Policy Optimization} (Masked PPO)~\cite{huang2022closer, schulman2017proximal}, implemented in the SB3-Contrib library on top of Stable-Baselines3.

\textbf{Policy architecture.}
The policy $\pi_\theta$ and value function $V_\phi$ share no parameters and are each implemented as a multi-layer perceptron (MLP) with two hidden layers of 64 units and $\tanh$ activations.
The policy MLP outputs logits $z_a$ for each action $a \in \mathcal{A}$.
Before softmax normalization, invalid actions are masked by setting their logits to $-\infty$:
\begin{equation}
    \pi_\theta(a | s_t) = \frac{m_t(a) \cdot \exp(z_a)}{\sum_{a'} m_t(a') \cdot \exp(z_{a'})},
\end{equation}
where $m_t(a) \in \{0, 1\}$ is the action mask.
This guarantees that the agent never proposes duplicate edges, eliminating the need for reward penalties or post-hoc rejection of invalid actions.

\textbf{PPO objective.}
Policy parameters are updated by maximizing the clipped surrogate objective:
\begin{equation}
    \mathcal{L}^{\mathrm{CLIP}}(\theta) = \hat{\mathbb{E}}_t \left[\min\left(r_t(\theta) \hat{A}_t,\ \mathrm{clip}(r_t(\theta), 1{-}\epsilon, 1{+}\epsilon) \hat{A}_t\right)\right],
    \label{eq:ppo_methods}
\end{equation}
where $r_t(\theta) = \pi_\theta(a_t | s_t) / \pi_{\theta_\text{old}}(a_t | s_t)$ is the probability ratio, $\hat{A}_t$ is the generalized advantage estimate (GAE)~\cite{schulman2015high} with $\gamma = 0.99$ and $\lambda = 0.95$, and $\epsilon = 0.2$ is the clipping parameter.
An entropy bonus $\mathcal{H}[\pi_\theta(\cdot | s_t)]$ weighted by a coefficient $c_\mathrm{ent}$ encourages exploration, which is important in the early phase when the agent has not yet discovered which edges are rewarding.

\textbf{Training loop.}
At each PPO update, the agent collects a rollout of $n_\text{steps} = 512$ environment transitions, then performs $n_\text{epochs}$ gradient passes over the data in minibatches of size $64$.
The value function is trained via mean squared error on the returns.
We found that reducing $n_\text{epochs}$ from $10$ to $5$ and the entropy coefficient from $0.1$ to $0.03$ improved convergence for larger instances (see Section~\ref{sec:results}).

\section{Experimental Setup}
\label{sec:experimental_setup}

\subsection{Problem Instances}
\label{sec:setup:instances}
We conduct our experiments on unweighted Erd\H{o}s--R\'{e}nyi random graphs $G(n, 0.5)$ with $n \in \{6, 8, 10\}$ nodes.
Each graph is generated with an edge probability of $0.5$ and verified to be connected.
All reported experiments use unweighted graphs; the framework supports weighted instances, but these are not included in the current evaluation.

\subsection{Training Configuration}
Table~\ref{tab:hyperparams} summarizes our training hyperparameters.
We train two configurations: one with tuned hyperparameters (ent\_coef$=0.03$, $n_{\text{epochs}}=5$) and a baseline with default values (ent\_coef$=0.1$, $n_{\text{epochs}}=10$).
Each configuration is run with 5 random seeds per graph size.

\begin{table}[t]
\centering
\caption{Hyperparameters for the RL agent and VQA optimizer.}
\label{tab:hyperparams}
\begin{tabular}{lcc}
\toprule
\textbf{Parameter} & \textbf{Tuned} & \textbf{Baseline} \\
\midrule
\multicolumn{3}{l}{\textit{RL Agent (Masked PPO)}} \\
Learning rate & $3 \times 10^{-4}$ & $3 \times 10^{-4}$ \\
Rollout length ($n_{\text{steps}}$) & 512 & 512 \\
Minibatch size & 64 & 64 \\
PPO epochs ($n_{\text{epochs}}$) & 5 & 10 \\
Discount ($\gamma$) & 0.99 & 0.99 \\
Clip range ($\epsilon$) & 0.2 & 0.2 \\
Entropy coef. & 0.03 & 0.1 \\
Network architecture & [64, 64] & [64, 64] \\
\midrule
\multicolumn{3}{l}{\textit{VQA Optimization}} \\
QAOA layers ($p$) & 1 & 1 \\
Optimizer steps (training) & 50 & 50 \\
Learning rate & 0.05 & 0.05 \\
\midrule
\multicolumn{3}{l}{\textit{Environment}} \\
Edge budget ($m_{\max}$) & $2n$ & $2n$ \\
Reward mode & sparse & sparse \\
Action masking & unique & unique \\
\midrule
\multicolumn{3}{l}{\textit{Training budget}} \\
6 qubits & 10{,}000 & 10{,}000 \\
8 qubits & 20{,}000 & 20{,}000 \\
10 qubits & 40{,}000 & 40{,}000 \\
\bottomrule
\end{tabular}
\end{table}

\subsection{Conducted Experiment Suites}
\label{sec:setup:suites}

We report results from two experiment suites with the following configurations:
\begin{itemize}
    \item \textbf{Baseline sparse suite:} seeds $\{1,\dots,5\}$ for $n \in \{6,8,10\}$, unique-action masking, sparse rewards, ent\_coef $0.1$, $n_{\text{epochs}}{=}10$, $p{=}1$, 50 optimization steps during training.
    \item \textbf{Tuned sparse suite:} seeds $\{1,\dots,5\}$ for $n \in \{6,8,10\}$, unique-action masking, sparse rewards, ent\_coef $0.03$, $n_{\text{epochs}}{=}5$, $p{=}1$, 50 optimization steps during training.
\end{itemize}

\subsection{Topology Comparison}
\label{sec:comparison}

To contextualize the RL-discovered topologies, we compare against several baselines.
Post-training comparisons are evaluated with multiple independent optimization restarts (seeds $\{1,\dots,5\}$), varying the number of optimization steps to study convergence behavior.
Each comparison includes multiple random draws per random-topology setting.

\begin{itemize}
    \item \textbf{Full}: all edges present in the problem graph (standard QAOA),
    \item \textbf{Linear}: nearest-neighbor chain $(0\text{--}1, 1\text{--}2, \ldots)$,
    \item \textbf{Ring}: linear chain plus closing edge,
    \item \textbf{Random $k\%$}: random subsets of $k\%$ of problem edges ($k \in \{25, 50, 75\}$),
    \item \textbf{Random RL-sized}: random subset with the same number of edges as the RL topology.
\end{itemize}

\subsection{Evaluation Metrics}
\label{sec:setup:metrics}

\begin{enumerate}
    \item \emph{Approximation Ratio} $\text{AR} = -\langle\psi(\bm{\theta}^*)|H_C|\psi(\bm{\theta}^*)\rangle / C^*$, benchmarked against brute-force optimal.
    \item \emph{Edge Count} $|\mathcal{T}|$, the number of entangling gates in the topology.
    \item \emph{Overlap Ratio} $|\mathcal{T} \cap E| / |\mathcal{T}|$, the fraction of RL-selected edges which are also present in the problem graph.
    \item \emph{Jaccard Similarity} $J(\mathcal{T}, E) = |\mathcal{T} \cap E| / |\mathcal{T} \cup E|$, a symmetric measure of topology alignment.
    \item \emph{Meyer--Wallach entanglement measure} $Q(\ket{\psi})$ as defined in Eq.~\ref{eq:mw}.
\end{enumerate}

\section{Results}
\label{sec:results}

We present results in five parts: computational cost, training dynamics and convergence, emergent alignment with the problem graph, topology comparison at different optimization budgets, and scaling behavior with deeper QAOA circuits.

\subsection{Computational Cost}
\label{sec:results:cost}
Each RL training run requires evaluating the VQA inner loop once per episode.
For $n{=}10$ with $40{,}000$ timesteps and an edge budget of $m_{\max}{=}20$, this corresponds to approximately $2{,}000$--$40{,}000$ episodes depending on when the agent triggers \texttt{STOP}, each involving $50$ gradient steps of QAOA optimization on the statevector simulator.
A single training run for $n{=}10$ completes in approximately $2$--$4$ hours on a standard CPU.\footnote{Timings are approximate and depend on hardware; our experiments used a single-threaded statevector simulator.}
By comparison, optimizing the full topology with $500$ gradient steps on the same instance takes only seconds---highlighting that the RL framework's value lies not in computational savings but in the structural insights it provides about the relationship between topology and trainability.

\subsection{Training Dynamics}
\label{sec:results:dynamics}

Table~\ref{tab:training_results} summarizes the training outcomes across all configurations.
The RL agent learns to construct topologies achieving approximation ratios of $0.72$--$0.81$ across problem sizes, using only $6$--$8$ edges out of $15$--$45$ possible pairs.
The tuned configuration consistently outperforms the baseline, particularly at $10$ qubits ($\text{AR} = 0.810$ vs.\ $0.747$).

\begin{table}[t]
\centering
\caption{Training results (mean $\pm$ std over $5$ seeds, last $20$ episodes). $|E|$ denotes problem graph edges; $|\mathcal{T}|$ denotes RL-selected edges.}
\label{tab:training_results}
\begin{tabular}{llcccc}
\toprule
\textbf{Config} & $n$ & $\text{AR}$ & $|\mathcal{T}|$ & $|E|$ & $Q$ \\
\midrule
\multirow{3}{*}{Tuned} & 6 & $.723 \pm .050$ & $5.8$ & $7.6$ & $.699$ \\
 & 8 & $.783 \pm .043$ & $6.9$ & $13.8$ & $.688$ \\
 & 10 & $.810 \pm .041$ & $8.0$ & $20.2$ & $.824$ \\
\midrule
\multirow{3}{*}{Baseline} & 6 & $.737 \pm .046$ & $5.8$ & $7.6$ & $.708$ \\
 & 8 & $.764 \pm .037$ & $7.1$ & $13.8$ & $.675$ \\
 & 10 & $.747 \pm .027$ & $8.7$ & $20.2$ & $.646$ \\
\bottomrule
\end{tabular}
\end{table}

\begin{figure}[t]
\centering
\includegraphics[width=\columnwidth]{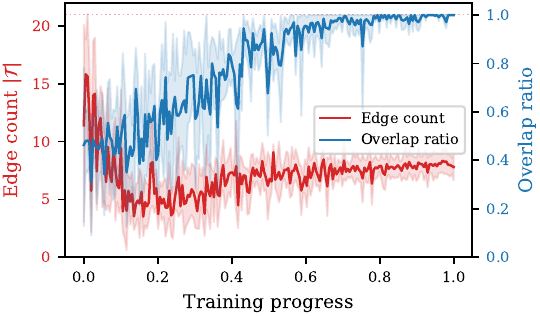}
\caption{Training dynamics for 10-qubit instances (mean $\pm$ std over 5 seeds). The edge count decreases and stabilizes around $8$ (out of $21$ problem edges), while the overlap ratio---the fraction of selected edges belonging to the problem graph---rises from $\sim0.5$ to $\sim1.0.$}
\label{fig:training_dynamics}
\end{figure}

To understand \emph{how} the agent converges, Table~\ref{tab:convergence} tracks three quantities across training deciles for the $10$-qubit tuned configuration (aggregated over $5$ seeds).
Three concurrent trends emerge (see also Fig.~\ref{fig:training_dynamics}): (i)~the agent initially places many edges ($\sim$10) before learning to reduce to $\sim$8; (ii)~the approximation ratio steadily improves from $0.66$ to $0.81$; and (iii)~the overlap ratio---the fraction of selected edges that belong to the problem graph---rises from $0.47$ to $0.99$.

\begin{table}[t]
\centering
\caption{Evolution of key metrics across training for 10-qubit tuned configuration (mean over 5 seeds). The agent simultaneously learns to select fewer edges, improve solution quality, and converge to problem-aligned topologies.}
\label{tab:convergence}
\begin{tabular}{rccc}
\toprule
\textbf{Decile} & $|\mathcal{T}|$ & $\text{AR}$ & \textbf{Overlap} \\
\midrule
10\% & 9.8 & 0.660 & 0.468 \\
20\% & 4.9 & 0.671 & 0.543 \\
30\% & 5.1 & 0.683 & 0.623 \\
40\% & 6.1 & 0.706 & 0.730 \\
50\% & 6.9 & 0.727 & 0.836 \\
60\% & 7.2 & 0.746 & 0.916 \\
70\% & 7.4 & 0.764 & 0.962 \\
80\% & 7.6 & 0.782 & 0.984 \\
90\% & 7.8 & 0.797 & 0.989 \\
100\% & 8.1 & 0.809 & 0.992 \\
\bottomrule
\end{tabular}
\end{table}

The early exploration phase (first $10$--$20$\%) is characterized by many edges (the agent has not yet learned when to stop) and low overlap (many non-problem edges are sampled).
As training progresses, the agent learns two complementary strategies: \emph{pruning}---selecting fewer edges per episode---and \emph{targeting}---concentrating selections on problem-graph edges.
By the final decile, the overlap ratio exceeds 0.99, indicating that virtually all selected edges are problem-relevant.
Crucially, these two behaviors are learned \emph{simultaneously}: the agent does not first learn the correct edges and then learn to stop, nor vice versa, but develops both skills in tandem through reward maximization.

\subsection{Emergent Problem-Graph Alignment}
\label{sec:results:alignment}

A central finding is that the RL agent, without any explicit knowledge of the problem graph, consistently discovers topologies that are \emph{strict subsets} of the problem graph edges.

Table~\ref{tab:alignment} quantifies this alignment.
For $8$- and $10$-qubit instances, the overlap ratio reaches $1.000 \pm 0.000$ in the tuned configuration---meaning \emph{every single edge} selected by the RL agent belongs to the problem graph.
This is remarkable given that the agent's observation space contains no explicit information about which edges are in the problem graph; it can only observe the edges it has placed and the resulting approximation ratio.

\begin{table}[t]
\centering
\caption{Topology alignment with the problem graph (mean $\pm$ std over 5 seeds, final topology). Overlap = $|\mathcal{T} \cap E|/|\mathcal{T}|$; Compression = $|\mathcal{T}|/|E|$.}
\label{tab:alignment}
\begin{tabular}{llccc}
\toprule
$n$ & \textbf{Overlap} & \textbf{Jaccard} & \textbf{Compression} \\
\midrule
6 & $.882 \pm .098$ & $.656 \pm .189$ & $76\%$ \\
 8 & $1.00 \pm .000$ & $.498 \pm .114$ & $48\%$ \\
 10 & $1.00 \pm .000$ & $.404 \pm .108$ & $39\%$ \\
\bottomrule
\end{tabular}
\end{table}

The decreasing Jaccard similarity with increasing $n$ reflects the growing compression: at 10 qubits, the agent uses only $\sim$39\% of the problem graph edges ($\sim$8 out of $\sim$20), yet all selected edges are problem-relevant.
This suggests that the variational optimization landscape inherently favors problem-aligned entanglement, and the RL agent discovers this alignment through reward maximization alone.

Examining the final topologies at the per-seed level reveals that the subset property is not merely a statistical artifact but holds for \emph{every individual run} in the 8- and 10-qubit tuned configuration: across all 10 seeds, every selected edge is a problem-graph edge, with zero exceptions.

\subsection{Topology Comparison at Different Optimization Budgets}
\label{sec:results:comparison}

A natural question is whether the RL topology's advantage is robust across different evaluation conditions.
Table~\ref{tab:comparison_budget} compares the RL topology against baselines at two optimization budgets for the 10-qubit tuned suite.

\begin{table}[t]
\centering
\caption{Topology comparison on 10-qubit instances (tuned config, 5 seeds). The RL topology dominates at low optimization budgets but the full topology recovers at high budgets.}
\label{tab:comparison_budget}
\begin{tabular}{lcc}
\toprule
\textbf{Topology} & \textbf{AR (50 opt.\ steps)} & \textbf{AR (500 opt.\ steps)} \\
\midrule
\textbf{RL (ours)} & $\bm{.808 \pm .038}$ & $.832 \pm .047$ \\
Full (problem graph) & $.676 \pm .009$ & $\bm{.860 \pm .047}$ \\
Random RL-sized & $.751 \pm .032$ & $.805 \pm .032$ \\
Random 75\% & $.723 \pm .043$ & $.849 \pm .034$ \\
Random 50\% & $.746 \pm .037$ & $.822 \pm .037$ \\
Random 25\% & $.721 \pm .030$ & $.769 \pm .035$ \\
Ring & $.688 \pm .031$ & $.775 \pm .035$ \\
Linear & $.687 \pm .039$ & $.765 \pm .043$ \\
\bottomrule
\end{tabular}
\end{table}

\begin{figure*}[t]
\centering
\includegraphics[width=\textwidth]{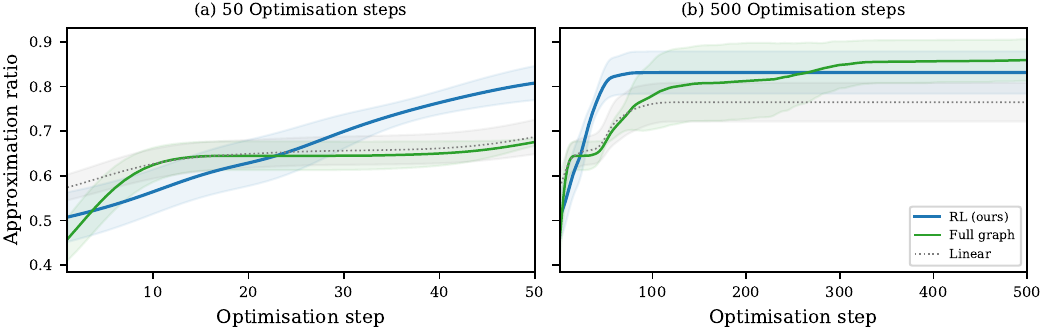}
\caption{QAOA optimization trajectories for different topologies on 10-qubit instances. (a)~At 50 steps, the RL topology converges fastest and reaches the highest AR. (b)~At 500 steps, the full graph topology eventually surpasses the RL topology, illustrating the trainability--expressibility trade-off.}
\label{fig:budget_comparison}
\end{figure*}

\begin{figure*}[t]
\centering
\includegraphics[width=\textwidth]{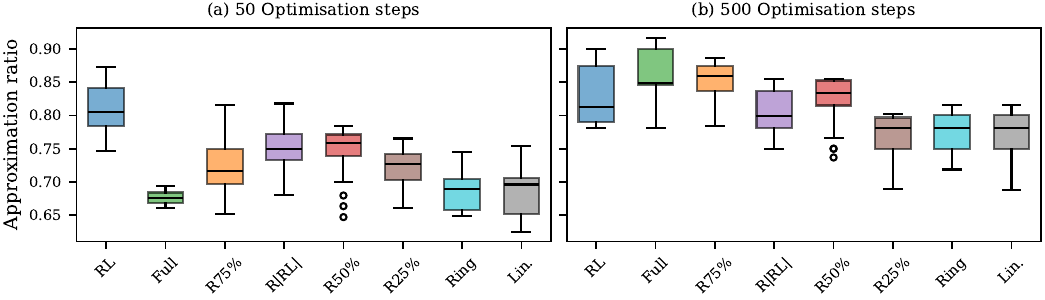}
\caption{Final approximation ratio distributions across topology types for 10-qubit instances. (a)~At 50 optimization steps, the RL topology clearly dominates. (b)~At 500 steps, the full topology achieves the best median AR.}
\label{fig:boxplot}
\end{figure*}

Figures~\ref{fig:budget_comparison} and~\ref{fig:boxplot} visualize these results.
At $50$ optimization steps---the same budget used during RL training---the learned topology achieves $\text{AR} = 0.808$, outperforming \emph{all} baselines by a wide margin.
The full graph topology, despite having the highest circuit expressibility, achieves only $\text{AR} = 0.676$, substantially worse.
This inversion occurs because the full topology introduces $\sim$20 IsingZZ parameters per layer, creating a high-dimensional optimization landscape that 50 gradient steps cannot adequately explore.
The RL topology, with only $\sim$8 edges, has fewer parameters and converges faster.

At 500 optimization steps, the full topology has sufficient budget to exploit its greater expressibility and overtakes the RL topology ($0.860$ vs.\ $0.832$).
However, the RL topology still outperforms all other baselines: it exceeds the random-RL-sized baseline by 2.7 percentage points ($0.832$ vs.\ $0.805$), confirming that edge \emph{selection} matters beyond edge \emph{count}.

This pattern---RL dominance at low budgets, full-graph recovery at high budgets---has practical relevance: on near-term quantum hardware, where circuit repetitions and coherence times limit the effective optimization budget, sparse problem-aligned topologies may offer better performance than full connectivity.

\subsection{Scaling with Circuit Depth}
\label{sec:results:depth}

The topologies reported above were discovered using $p=1$ QAOA layers, where the achievable approximation ratios are inherently limited.
A natural question is whether the RL-discovered topologies retain their advantage when used with deeper circuits.
To investigate, we evaluate each topology from the 10-qubit tuned suite with $p \in \{1, 2, 3\}$ QAOA layers and 300 optimization steps.

\begin{table}[t]
\centering
\caption{Approximation ratios for different topologies as a function of QAOA depth $p$ (10-qubit tuned suite, 300 optimization steps). The RL topology, discovered at $p{=}1$, does not retain its advantage at deeper circuits: the full topology reaches AR${=}1.000$ at $p{=}2$.}
\label{tab:depth_scaling}
\begin{tabular}{lccc}
\toprule
\textbf{Topology} & $p=1$ & $p=2$ & $p=3$ \\
\midrule
RL (ours) & $.801$ & $.913$ & $.916$ \\
Full (problem graph) & $.820$ & $\bm{1.000}$ & $.984$ \\
Random 75\% & $.843$ & $.966$ & $\bm{.997}$ \\
Random RL-sized & $.801$ & $.855$ & $.857$ \\
Random 50\% & $.831$ & $.901$ & $.960$ \\
Ring & $.792$ & $.829$ & $.945$ \\
Linear & $.792$ & $.803$ & $.893$ \\
Random 25\% & $.785$ & $.795$ & $.827$ \\
\bottomrule
\end{tabular}
\end{table}

Table~\ref{tab:depth_scaling} reveals a clear trend: as circuit depth increases, topologies with more edges benefit disproportionately.
At $p{=}2$, the full topology achieves $\text{AR} = 1.000$ while the RL topology reaches only $0.913$.
The random 75\% subset also approaches near-optimal performance ($0.966$), confirming that expressibility, not edge selection, is the dominant factor at sufficient depth.

The RL topology, with only $\sim$8 edges, plateaus around $\text{AR} \approx 0.91$ regardless of depth ($p{=}2$ and $p{=}3$ yield nearly identical results).

This result reinforces the trainability--expressibility trade-off identified in Section~\ref{sec:results:comparison}: the RL topology's advantage is specific to the low-budget regime where trainability dominates.
With sufficient optimization budget \emph{and} circuit depth, denser topologies---particularly the full problem graph---are strictly superior.

\section{Discussion}
\label{sec:discussion}

\subsection{Why Subsets?}
The emergent subset property can be understood through the structure of the MaxCut Hamiltonian.
Each term $w_{ij}(Z_i Z_j - I)/2$ creates correlations between qubits $i$ and $j$.
An IsingZZ gate on an edge $(i,j) \in E$ directly addresses the corresponding Hamiltonian term, enabling the optimizer to tune the correlation between these qubits.
Conversely, an IsingZZ gate on a non-problem edge $(i,j) \notin E$ introduces correlations that do not correspond to any Hamiltonian term, hindering optimization by coupling qubits that should be independently assigned.

The subset (rather than full graph) nature of the learned topologies suggests that \emph{not all} Hamiltonian terms require direct entangling gates.
Some correlations can be mediated indirectly through paths in the topology graph, as long as a sufficient ``backbone'' of problem-aligned edges is present.

We note that this argument is qualitative; a rigorous proof that non-problem edges strictly impair optimization would require a landscape analysis beyond the scope of this work.
However, the empirical evidence is consistent: across all 10 seeds at $n{=}8$ and $n{=}10$, the converged topologies suggest strong negative reward pressure against such edges.

\subsection{The Trainability--Expressibility Trade-off}

Our results across optimization budgets (Table~\ref{tab:comparison_budget}) and circuit depths (Table~\ref{tab:depth_scaling}) illuminate a fundamental trade-off in variational quantum circuit design.
The full topology maximizes \emph{expressibility}---the set of reachable quantum states---but at the cost of \emph{trainability}: more parameters require more optimization steps to converge.
The RL-learned topology operates here in a favorable regime for fast convergence, but hits an expressibility ceiling when more compute is available.

Concretely, the RL topology dominates at low optimization budgets (50 steps: $0.808$ vs.\ $0.676$ for full) but is overtaken when the optimizer has sufficient budget (500 steps: $0.832$ vs.\ $0.860$) or depth ($p{=}2$: $0.913$ vs.\ $1.000$).
This suggests that, \emph{in the regime studied} ($p{=}1$, budgets up to $500$ steps), entanglement topology matters primarily for convergence speed rather than asymptotic solution quality.
Whether this conclusion extends to higher depths or larger instances remains an open question.

This finding has practical implications in two directions.
First, on near-term quantum hardware where coherence times and noise constrain effective circuit depth, sparse problem-aligned topologies may offer practical advantages.
Second, it motivates investigating whether RL agents trained at higher depths ($p > 1$) would learn denser topologies that better balance the trade-off.

\subsection{Hardware Topology Constraints}

On real quantum hardware, full connectivity is typically unavailable---devices have fixed coupling maps with limited connectivity.
Our framework could be adapted to discover optimal topologies \emph{within} hardware constraints, reframing the question from ``RL vs.\ full'' to ``RL-selected subset vs.\ random/heuristic subset of available connections.''
In this setting, the RL agent's ability to identify problem-aligned edges from a restricted set may prove more valuable than in the unconstrained simulation studied here.

\subsection{Limitations}

Our study has several limitations that should be considered when interpreting the results.

\textbf{Scale and statistical power.}
All experiments use at most $10$~qubits, where brute-force optimal solutions are available.
While the strengthening trend from $n{=}6$ to $n{=}10$ (increasing overlap, decreasing compression ratio) is suggestive, we cannot confirm that the subset property persists at larger scales where classical simulation becomes infeasible.
Results are averaged over $5$~seeds without formal significance tests which are not reasonable for this sample size; the observed differences (e.g., $0.808 \pm 0.038$ vs.\ $0.751 \pm 0.032$) should be interpreted with this limited statistical power in mind.

\textbf{Instance specificity.}
Each training run uses a single fixed graph, and each qubit count is evaluated on only one graph topology class (Erd\H{o}s--R\'{e}nyi with $p{=}0.5$).
The agent therefore learns topology preferences specific to each training instance; whether these preferences generalize to unseen instances of the same class, or to other graph families (e.g., regular graphs, planar graphs, or real-world networks), would require meta-learning or transfer-learning extensions.

\textbf{Budget and depth conditioning.}
The RL agent is trained exclusively with $p{=}1$ QAOA layers and $50$ VQA optimization steps.
The discovered topologies are therefore optimized for fast convergence under this specific regime.
As the budget-crossover (Table~\ref{tab:comparison_budget}) and depth-scaling (Table~\ref{tab:depth_scaling}) experiments show, the advantage does not transfer to higher budgets or deeper circuits, suggesting that topology preferences are jointly conditioned on optimization budget and circuit depth.

\textbf{Noise-free simulation.}
We use statevector simulation without hardware noise.
Under realistic noise conditions, sparse topologies may benefit further from reduced error accumulation in two-qubit gates, but this hypothesis remains untested.

\section{Conclusion and Future Work}
\label{sec:conclusion}

We have presented an RL framework for discovering entanglement topologies in variational quantum circuits for combinatorial optimization.
Our experiments reveal two key findings.

First, RL agents guided solely by the approximation ratio consistently discover topologies that are strict subsets of the problem graph---selecting only 30--50\% of problem edges with overlap ratios approaching 1.0.
This emergent alignment, achieved without explicit problem-graph information in the agent's observation, demonstrates that the variational optimization landscape encodes structural information about the problem Hamiltonian.

Second, the RL topology's advantage is \emph{budget-dependent}: at low optimization budgets (50 steps), the sparse learned topology outperforms the full graph by over 13 percentage points, but with sufficient budget (500 steps) or circuit depth ($p \geq 2$), denser topologies recover and eventually surpass it.
In the regime studied with short optimization cycles, this reveals that entanglement topology governs primarily convergence speed rather than asymptotic solution quality.

These findings have practical implications for near-term quantum computing, where limited coherence times constrain effective optimization budgets, and for hardware-constrained settings where full connectivity is unavailable.
Future work will investigate RL training at higher circuit depths, hardware-aware topology discovery under coupling-map constraints, GNN-based policy networks for cross-instance generalization, and validation on noisy quantum hardware.

\section*{Acknowledgment}
\label{sec:acknowledgment}
This paper was partially funded by the German Federal Ministry of Education and Research through the funding program “quantum technologies -- from basic research to market” (contract number: 13N16196). Generative AI was utilized to generate sections of this Work, including text, tables, graphs, code, citations, etc. 


\bibliographystyle{IEEEtran}
\bibliography{references}

\end{document}